\documentclass[10pt,twoside]{article}
\usepackage{graphicx}
\usepackage{amsmath}
\usepackage{Latex-document}

\newcommand{\be}{\begin{equation}}
\newcommand{\ee}{\end{equation}}
\newcommand{\beq}{\begin{eqnarray}}
\newcommand{\eeq}{\end{eqnarray}}
\newcommand{\bt}{\begin{theorem}}
\newcommand{\et}{\end{theorem}}
\newcommand{\bl}{\begin{lemma}}
\newcommand{\el}{\end{lemma}}
\newcommand{\bc}{\begin{corollary}}
\newcommand{\ec}{\end{corollary}}
\newcommand{\bp}{\begin{prop}}
\newcommand{\ep}{\end{prop}}
\newcommand{\ba}{\begin{array}}
\newcommand{\ea}{\end{array}}

\newcommand{\la}{\label}
\newcommand{\ci}{\cite}

\newtheorem{theorem}{Theorem}
\newtheorem{lemma}[theorem]{Lemma}
\newtheorem{corollary}[theorem]{Corollary}
\newtheorem{prop}[theorem]{Proposition}

\newcommand{\eps}{\epsilon}

\newcommand{\ga}{\gamma}

\newcommand{\om}{\omega}

\newcommand{\lb}{\lambda}

\newcommand{\bi}{\bibitem}

\newfont{\msbm}{msbm10 scaled\magstep1}
\newfont{\msbms}{msbm7 scaled\magstep1} 

\newcommand{\RR}{\mbox{$\mbox{\msbm R}$}}
\newcommand{\ZZ}{\mbox{$\mbox{\msbm Z}$}}
\newcommand{\QQ}{\mbox{$\mbox{\msbm Q}$}}
\newcommand{\TT}{\mbox{$\mbox{\msbm T}$}}

\title{\bf Nonperturbative Localization\thanks{This research was partially supported by NSF grant DMS-0070755.}\vskip 6mm}

\author{S. Jitomirskaya\vspace*{-0.5cm}\thanks{Department of Mathematics, University of California at Irvine,
CA 92697, USA. E-mail:szhitomi@math.uci.edu}}
\date{\vspace{-8mm}}

\begin{document}

\maketitle

\thispagestyle{first} \setcounter{page}{445}

\begin{abstract}

\vskip 3mm

 Study of fine spectral properties of quasiperiodic and similar
 discrete \linebreak
Schrodinger operators involves dealing with problems caused by small
denominators, and until recently was only possible using perturbative
methods, requiring certain small parameters and complicated KAM-type
schemes. We review the recently developed nonperturbative methods for
such study which lead to stronger results and are significantly simpler.
Numerous applications mainly due to J. Bourgain, M. Goldstein, W. Schlag,
and the author are also discussed.

\vskip 4.5mm

\end{abstract}

\vskip 12mm

\section{Introduction}\label{section 1}\setzero

\vskip-5mm \hspace{5mm}

Consider an operator
acting on on $\ell^2(\ZZ^d)$ defined by
\begin{equation}
\la{op1} H_{\lambda}= \Delta +\lambda V,
\end{equation}
where $\Delta$ is the lattice tight-binding Laplacian
$$
\Delta(n,m)=\begin{cases} 1, & \mbox{dist}(n,m)=1, \cr 0, & \mbox {otherwise}, \end{cases}
$$
and $V(n,m)=V_n\delta(n,m)$ is a potential given by $V_n=f(T_1^{n_1}\cdots T_d^{n_d}\theta),\;\theta \in \TT^b,$
where $T_i\theta=\theta+\om_i,$ and $\om$ is an incommensurate vector. In certain cases
$\Delta$ may also be replaced by a long-range Laplacian. Replacing $T_i$'s with other commuting ergodic
transformations would give a general framework of ergodic
Schr\"odinger operators (\ci{cfks}; see Sec. \ref{kicked} for an example of this
kind) but we will mostly focus on the quasiperiodic (QP)
operators that
%
%
have been intensively studied  in Physics and Mathematics
literature.  For another review of some recent developments in this area see \ci{brev}.
The questions of interest are the nature and structure of the
spectrum, behaviour of the eigenfunctions, and particularly the  quantum
dynamics: properties of the time evolution $\Psi_t=e^{itH}\Psi_0$ of
an initially localized wave packet $\Psi_0.$

Of particular importance is the phenomenon of Anderson localization (AL) which is usually referred to the property of having
pure point spectrum with exponentially decaying eigenfunctions.
A somewhat stronger property of dynamical localization (see Sec. \ref{secdl})
 indicates  the insulator behavior,
while ballistic transport, which for $d=1$ follows from the
absolutely continuous (ac) spectrum, indicates the metallic behavior.

Operators with ergodic potentials always have spectra (and pure point (p.p.)
spectra, understood as closures of
the set of eigenvalues) constant for a.e. realization of the
potential. Moreover, the p.p. spectrum of operators with
ergodic potentials never contains
isolated eigenvalues, so p.p. spectrum in such models is dense
in a certain closed set.
An easy example of an operator with dense pure point spectrum is
 $H_{\infty}$ which is operator (\ref{op1}) with $\lambda^{-1}=0,$ or pure diagonal. It has a
complete set of eigenfunctions, characteristic functions of lattice
points, with eigenvalues $V_j.$ $H_{\lb}$ may be viewed
as a perturbation of  $H_{\infty}$ for small $\lambda^{-1}.$
 However, since $V_j$ are dense, small denominators $(V_i-V_j)^{-1}$
 make any perturbation theory difficult, e.g. requiring
 intricate KAM-type schemes.

The probabilistic KAM-type scheme was developed by Fr\"ohlich and
Spencer \ci{fs} for random potentials ($V_n$ are i.i.d.r.v.'s) in the multi-dimensional case,
and is called multi-scale analysis. It was significantly modified,
improved, and widely applied in the later years by a number of
authors, most notably \ci{dk}. An alternative method for random
localization was found by Aizenman-Molchanov \ci{am} and later further
developed by Aizenman and coauthors. While still requiring certain
large parameters this method relies on direct estimates of the Green's
function rather than a step-by-step perturbation scheme.

For QP potentials none of the above methods work, as, among
other reasons, they do not allow rank-one perturbations, nor
Wegner-type estimates. The situation here is more difficult and the
theory  is far less developed than for the random case. With a few
exceptions the results are confined to the 1D
 case, and also 1-frequency case ($b=1$) has been much better
developed than that of higher frequencies.

One might expect that $H_{\lb}$ with $\lb$ small can be treated as
a perturbation of $H_0=\Delta,$ and therefore have ac spectrum. It is
not the case though for random potentials
in $d=1,$ where AL holds for all $\lb.$ Same is
expected for random potentials in $d=2$ (but not higher). Moreover, in
1D case there is strong evidence (numerical, analytical,
as well as rigorous \ci{b2}) that even models with very mild
stochasticity in the underlying dynamics have point spectrum for all
values of $\lb$ like in the random case
(e.g. $V_n=\lb f(n^{\sigma}\alpha+\theta)$ for any $\sigma>1).$
At the same time, for QP potentials one can in many cases
show ac spectrum for $\lb$ small as well as pure
point spectrum for $\lb$ large (see below), and therefore there is a
metal insulator transition in the coupling constant. It is an
interesting question whether quasiperiodic potentials are the only
ones with metal-insulator transition in 1D.

\section{Perturbative vs nonperturbative}\setzero

\vskip-5mm \hspace{5mm}

It is probably
fair to say that much of the theory of qusiperiodic operators has
been first developed around the almost Mathieu operator, which is
\begin{equation}\la{am}
H_{\lb,\om,\theta}=\Delta + \lb f(\theta + n\om)
\end{equation}
 acting on $\ell^2(\ZZ),$
with $f:\TT\to\TT;\; f(\theta)= \cos (2\pi \theta).$
The first KAM-type approaches, in both
large and small coupling regimes, were developed for this or similar
models \ci{ds,blt}.
The perturbative
proofs of complete Anderson localization for $\lb>\lb(\om)$ large are
due to Sinai \ci{s} and Fr\"ohlich-Spencer-Wittwer \ci{fsw}, and both
applied to $\cos$-type $f.$ For $\lb$ small
Chulaevsky-Delyon \ci{cd} proved pure ac spectrum
using duality and the construction of Sinai \ci{s}.
Elliasson (see \ci{eli} for a review) developed  alternative
KAM-type arguments for both large and small $\lb$ for the case of real-analytic
(actually, somewhat more general) class of $f$ in (\ref{am}).

The common feature of the perturbative {\it approaches} above is that,
besides all of them being rather intricate multi-step procedures, they rely
extensively on
eigenvalue and eigenfunction parametrization and perturbation
arguments.

The common feature of the perturbative {\it results} in the quasiperiodic
setting is that they provide no explicit
estimates on how large (or small) the parameter $\lb$ should be, and,
more importantly, $\lb$ clearly depends on $\omega$ at least through
the constants in the Diophantine characterization of $\om.$

In contrast, the nonperturbative results
allow effective (in many cases even optimal) and, most importantly,
independent of $\om,$ estimates on $\lb.$ We will take the latter
property (uniform in $\om$ estimates on $\lb$) as a definition of a
NP result.

Recently developed nonperturbative methods are also quite different from the perturbative
ones, in that they do not employ multi-scale schemes: usually only a
few (from one to three) sufficiently large scales are involved, do not
use the eigenvalue parametrization, and rely instead on direct
estimates of the Green's function. They are also significantly less
involved, technically. One may think that in these latter respects they
resemble the Aizenman-Molchanov method for random localization. It is,
however, a superficial similarity, as, on the technical side, they are
still closer to and do borrow certain ideas from \ci{fs,dk}.

 Several results
that
satisfy our definition of nonperturbative appeared prior to the recent
developments, and were all related to the almost Mathieu operator (see \ci{j} for a review).
In \ci{j1,j2} AL was proved for $\lb >15,$ and
existence of p.p. component for $\lb > 2.$ The
latter papers, while introducing some of the ingredients of the recent
nonperturbative methods, did not take advantage of the positivity of
the Lyapunov exponents which proved very important later.

\section{Lyapunov exponents}\setzero

\vskip-5mm \hspace{5mm}

Here for simplicity we consider the quasiperiodic case,
although the definition of the Lyapunov exponents and some of the
mentioned facts apply more generally to the 1D ergodic case.

For an energy $E\in \RR$ the Lyapunov exponent $\ga(E)$ is defined as
\begin{equation}
\ga(E)=\lim _{n \to \infty} {\int_0^1\ln \|M_k(\theta,E)\|d\theta\over k},
\end{equation}
where
$$
M_k(\theta,E) =\prod_{n=k-1}^0\begin{pmatrix} E-\lb f(\om n+\theta)& -1 \cr 1 & 0 \end{pmatrix}
$$
is the $k-$step transfer-matrix for the eigenvalue equation $
H\Psi=E\Psi.
$

We will be
interested in the regime when Lyapunov exponents are positive for all
energies in a certain interval intersecting the spectrum.
 It is well known and fairly easy to see that
if  this condition holds for all $E\in \RR,$
 there is no ac component in
the spectrum for a.e. $\theta$ (it is actually true for all $\theta$ \ci{ls}).  Positivity of Lyapunov exponents, however, does not
imply exponential decay of eigenfunctions (in particular, not for the
Liouville $\om$ \ci{as} nor for the resonant $\theta \in \TT^b$ \ci{js}).

NP methods, at least in their original form, stem to a
large extent
from estimates involving the Lyapunov exponents and exploiting their
positivity.

The general theme of the results on
positivity of $\gamma(E)$, as suggested by perturbation arguments, is that the
Lyapunov exponents are positive for large $\lb.$ This was first
established by Aubry-Andre \ci{aa} for the almost Mathieu operator
with $\lb>2.$ Their proof was made rigorous in \ci{as}.
 Another proof, exploiting the
subharmonicity, was given by Herman \ci{her}, and applied to
trigonometric polynomials $f.$ The lower bound in \ci{her} was in terms of the highest
coefficient of the  trigonometric polynomial and therefore
this  did not easily extend to the real analytic case. All the subsequent proofs, however, were also based on subharmonicity.
Sorets-Spencer
\ci{ss} proved  that for nonconstant real analytic potentials $v$ on $\TT$ ($b=1$) one
has
$\gamma(E) > \frac{1}{2} \ln \lb
$
for $\lb > \lb (v)$ and all irrational $\om.$
Another proof was given in \ci{bg}, where this  was also
extended to the multi-frequency case ($b>1$) with, however, the
estimate on $\lb$ dependent on the Diophantine condition on $\om.$
Finally, Bourgain \ci{b02} proved that Lyapunov exponents are
continuous in $\omega$ at every incommensurate $\omega$ (for $b>1;$
for $b=1$ this was previously established in \ci{bj3}), and that led
to the following Theorem which is the strongest result in this general
context
up to date:

\bt {\rm \ci{b02}} \la{lyapbourg} Let $f$ be a nonconstant real analytic function on $\TT^b,$ and $H$ given by
(\ref{op1}). Then, for $\lb>\lb(f),$ we have $ \gamma(E) > \frac{1}{2} \ln \lb $ for all $E$ and all
incommensurate vectors $\om.$ \et
\subsection{Corollaries of positive Lyapunov exponents}
\paragraph {The almost Mathieu operator}
On one hand the almost Mathieu operator, while simple-looking, seems to represent most of the
nontrivial properties expected to be encountered in the more general
case. On the other hand it has a very special feature: the duality
(essentially a Fourier) transform maps $H_{\lb}$ to $H_{4/
\lb},$ hence $\lb=2$ is the self-dual point.
Aubry-Andre \ci{aa} conjectured that for this model, for irrational
$\om$  a sharp
metal-insulator transition in the coupling constant $\lb$ occurs at
the critical value of coupling $\lb = 2:$ the
spectrum is pure point for $\lb >2$ and pure ac
for $\lb < 2.$ A second, related, conjecture was that the dual of ac
spectrum is pure point and vice versa. Both conjectures were modified based on the results of \ci{as,l,js}.
%
 The first modified
conjecture stated pure point spectrum for Diophantine $\om$
and a.e. $\theta$ for $\lb >2$ and pure ac  spectrum
for $\lb < 2$ for all $\om,\theta.$ As for the duality, the question,
after some prior developments, was resolved in \ci{gjls} where it was
shown that the dual of point spectrum is ac spectrum (the proof
applied in a more
general context), and it was used (together with
\ci{l1,hs}) to prove
that the spectrum is purely singular continuous at $\lb=2$ for a.e. $\om,\theta.$

As with the KAM methods, the almost Mathieu operator was the first model where the positivity of Lyapunov
exponents was effectively exploited:

\bt {\rm \ci{j}} Suppose $\om$ is Diophantine and $\ga(E,\om)>0$ for all $E\in [E_1,E_2].$ Then the almost Mathieu
operator
has Anderson localization in $[E_1,E_2]$
for
a.e. $\theta.$
\et

The condition on $\theta$ in \ci{j} was actually explicit (arithmetic)
and close to optimal.
This, combined with the mentioned results on the Lyapunov exponents
for the almost Mathieu operator \ci{her} and duality \ci{gjls}
led to the following corollary:

\bc The almost Mathieu operator $H_{\om,\lb,\theta}$ has
\begin{enumerate}
\item[$1^o$] \ci{j} for $\lb >2,$  Diophantine $\om \in \RR$ and almost every $ \theta \in \RR,$ only pure point  spectrum with
exponentially  decaying eigenfunctions.
\item [$2^{o}$] \ci{gjls} for $\lb=2,\;$ and a.e. $\om,\theta \in \RR$  purely singular-continuous  spectrum.
\item  [$3^{o}$] \ci{j,gjls}
 for $\lb <2,\;$Diophantine $\om \in \RR$ and a.e. $ \theta \in
\RR,$
 purely ac  spectrum.
\end{enumerate}
\ec

Precise arithmetic descriptions of $\om,\theta$ are available. Thus the Aubry-Andre conjecture is settled at least
for almost all $\om,\theta.$ One should mention, however, that while $1^o$ is almost optimal, both $2^o$ and $3^o$
are expected to hold for all $\theta$ and all $\om \not \in \QQ,$ and such extension remains a challenging problem
(see \ci{sim}).

The method in \ci{j}, while so far the only nonperturbative available allowing
precise arithmetic conditions, uses some specific properties of the
cosine. It extends to certain other but rather limited situations. A
much more robust method was developed by Bourgain-Goldstein \ci{bg},
which allowed them to extend (a measure-theoretic version of) the
above result to the general real analytic as well as the
multi-frequency case. Note, that essentially no results were
previously available for the multifrequency case, even perturbative.

\bt {\rm \ci{bg}} \la{thbg} Let $f$ be  non-constant real analytic on $\TT^b$ and $H$ given by (\ref{am}). Suppose
$\ga(E,\om)>0$ for all $E\in [E_1,E_2]$ and a.e. $\om \in \TT^b.$ Then for any $\theta,$ $H$ has Anderson
localization in $[E_1,E_2]$ for a.e. $\om.$
\et

Combining this with Theorem \ref{lyapbourg} one obtains \ci{b02} that for $\lb > \lb(f),$ $H$ as above
 satisfies Anderson localization for a.e. $\om.$

One very important ingredient of the method of \ci{bg} is the theory
of semi-algebraic sets that allows one to obtain polynomial algebraic
complexity bounds for certain ``exceptional'' sets. Combined
with measure estimates coming from the large deviation analysis of
${1\over n}\ln ||M_n(\theta)||$ (using subharmonic function theory
and involving approximate Lyapunov exponents), this theory provides necessary information on the geometric structure of those exceptional
sets. Such algebraic complexity bounds also exist for the almost
Mathieu operator \ci{j} and are actually
sharp  albeit  trivial in
this case due to the specific nature of the cosine.

\section{Without Lyapunov exponents}\setzero

\vskip-5mm \hspace{5mm}

While having led to significant advances, Lyapunov exponents have
obvious limitations, as any method, based on them, is restricted to 1D
nearest neighbor Laplacians. It turns out that the above methods can be
extended to obtain NP results in certain
quasi-1D situations where Lyapunov exponents do not
exist.

For the next Theorem let $H$ be an operator (\ref{op1}) defined on $\ell^2(S)$ where
$S=\ZZ\times S_0,$ is a strip. $S_0$ here is a finite set with a metric,
and $\mbox {dist} ((n,s),(n',s'))=|n-n'|+\mbox {dist}(s,s').$ Let
$V_{(n,s)}=f_s(\theta +n\om),\;\theta \in \TT.$

\bt {\rm \ci{bj}} \la{strip} Assume $f_s,\;s\in S_0,$ are non-constant real analytic functions on $\TT.$ Then for
any $\theta \in \TT $ and $\lb>\lb(f_s),$ operator $H$ has AL for a.e. $\om.$ \et

The following nonperturbative Theorem deals with the case of small coupling:

\bt \la{longr}{\rm \ci{bj2}} Let $H$ be an operator (\ref{am}), where $f$ is real analytic on $\TT$ and $\om$ is
Diophantine. Then, for $\lb < \lb(f),$ $H$ has purely ac spectrum for a.e. $\theta.$ \et

We note that an analogue of this Theorem does not hold in the
multi-frequency case (see next section).
Theorem
\ref{longr} is a result on non-perturbative localization in disguise
as it was obtained using duality \ci{gjls} from a localization Theorem
for a dual model which has in general a long-range Laplacian and
was in turn obtained by an extension of the method of
\ci{j}. A certain measure-theoretic version of it by the method of \ci{bg} allowing
non-local Laplacians but leading only to continuous spectrum is also
available \ci{bourglect}.
Theorem \ref{strip} was obtained by an extension of the method of
\ci{bg}. Both Theorems above rely on large deviations for the
quantities of the form ${1 \over n} \ln |\det(H-E)_{\Lambda}|$
and path-determinant expansion for the matrix elements of the
resolvent \ci{bj2}. The methods developed in \ci{bj2} apply also to
certain other situations with long-range Laplacians, for example the
kicked rotor model (see Sec. \ref{kicked}).

\section {Multidimensional case: \boldmath$d>1$}

\vskip-5mm \hspace{5mm}

As mentined above, there are very few results in the multidimensional
lattice case ($d>1$).
Essentially, the only result that existed before the new developments was a perturbative Theorem - an extension by
Chulaevsky-Dinaburg
\ci{cdi}  of Sinai's \ci{s} method to the case of operator (\ref{op1}) on $\ell^2(\ZZ^d)$ with
$V_n=\lb f(n\cdot\om),$
$\om \in \RR^d,$ where $f$ is a
$\cos$-type function on $\TT.$ Recently, Bourgain \ci{bourglect} obtained this result for real analytic $f$ by a nonperturbative
method. Note that since
$b=1,$ this avoids most serious difficulties and is therefore significantly simpler than the general multi-dimensional case.

\bt {\rm \ci{cdi,bourglect}} \la{Thcdi} For any $\eps >0$ there is $\lb(f,\eps),$ and, for $\lb > \lb (f,\eps),$
$\Omega(\lb,f) \subset \TT^d$ with $\mbox{mes} (\Omega) <\eps,$ so that for $\omega \notin \Omega,$ operator
(\ref{op1}) with $V_n$ as above has Anderson localization. \et

This should be confronted with the following Theorem of Bourgain
\ci{b2}

\bt {\rm \ci{b2}} Let $d=2$ and $f(\theta)=\cos 2\pi \theta$ in $H=H_{\om}$ defined as above. Then for any $\lb$
measure of $\om$ s.t. $H_{\om}$ has some continuous spectrum is positive. \et

Therefore for large $\lb$ there will be both $\om$ with complete localization
as well as those  with at least some continuous spectrum. This shows that nonperturbative {\it results} do not hold in general
in the multi-dimensional case!

A similar (in fact, dual) situation is observed for 1D multi-frequency ($d=1;\;b>1$)
case at small disorder.
One has, by duality:

\bt {\rm \ci{cdi,cd}} Let $H$ be given by (\ref{am}) with $\theta,\om \in \TT^b$ and $f$ real analytic on $\TT^b.$
Then for any  $\eps >0$ there is $\lb(f,\eps)$ s.t. for $\lb < \lb (f,\eps)$ there is $\Omega(\lb,f) \subset
\TT^b$ with $\mbox{mes} (\Omega) <\eps$ so that for $\omega \notin \Omega,\,$ $H$ has purely ac spectrum. \et

\bt {\rm \ci{b2}} Let $d=1,b=2$ and $f$ be a trigonometric polynomial on $\TT^2$ with a non-degenerate maximum.
Then for any $\lb$ measure of $\om$ s.t. $H_{\om}$ has some point spectrum, dense in a set of positive measure, is
positive. \et

Therefore, unlike the $b=1$ case (see Theorem \ref{longr}), nonperturbative $\it results$ do not
hold for absolutely continuous spectrum at small disorder.

\section{Perturbative results by NP methods}

\vskip-5mm \hspace{5mm}

While the above demonstrates the limitations of the NP
results, the nonperturbative {\it methods} have been applied to
significantly simplify the proofs and obtain new perturbative results
that previously have been completely beyond reach.

We refer the reader for a description of many such applications to
\ci{brev,bourglect}. In particular, new results on the construction of
QP solutions in Melnikov problems and nonlinear PDE's, obtained by using
certain ideas developed for NP quasi-periodic
localization (e.g. the theory of semi-algebraic sets) are presented
there.

We will only mention here a theorem by Bourgain-Goldstein-Schlag that
is the only one so far treating a ``true" $d>1$ situation. Note that here $d=2,$ and the reasons why it
has not yet been extended to higher dimensions are not just technical, but conceptual (there are certain purely arithmetic
difficulties).

\bt {\rm \ci{bgs}} \la{d=2} Let $d=b=2$ and $f$ be real analytic on $\TT^2$ such that all functions
$f(\theta_1,\cdot),\,f(\cdot,\theta_2),\;(\theta_1,\theta_2)\in \TT^2$ are nonconstant. Then for any $\eps >0$
there is $\lb(f,\eps)$ s.t. for $\lb > \lb (f,\eps)$ there is $\Omega(\lb,f) \subset \TT^d$ with $\mbox{mes}
(\Omega) <\eps$ so that for $\omega \notin \Omega$ operator (\ref{op1}) with $V_n=\lb f(n_1\om_1,n_2\om_2)$  has
Anderson localization. \et

\section{Dynamical localization} \la{secdl}

\vskip-5mm \hspace{5mm}

Anderson localization
 does not in itself guarantee absense of
quantum transport, or nonspread of an initially localized wave packet, as characterised,
e.g., by boundedness in time of moments of the position operator
\ci{djls} (\ci{jss} for an example of physical model with coexistence of exponential
localization and quantum transport). Considering for simplicity the second moment
$$\langle x^2\rangle_T = \frac{1}{ T}\int_0^T\displaystyle\sum_{n}|\Psi_t(n)|^2n^2 dt, $$
we will say that $H$ exhibits dynamical localization (DL) if $\langle
x^2\rangle_T < \mbox{Const}.$ We will say that the family $\{H_{\theta}\}_{\theta \in \TT^b}$ exhibits strong DL
if
$\int_{\TT^b}d\theta\displaystyle\sup_t \langle x^2\rangle_t < \mbox{Const}.$ We note that the results mentioned below will hold
with more restrictive definitions of DL (involving the higher moments of the position operator) as well. DL
implies p.p. spectrum by RAGE theorem (see,
e.g. \ci{cfks}), so it is a strictly stronger notion.

It turns out that nonperturbative methods allow for such dynamical upgrades as well.
For the almost Mathieu operator we have

\bt {\rm \ci{gj}} \la{thgj} For $\lb>2$ and Diophantine $\om$ (as in \ci{j}) strong DL holds. \et

While proved in \ci{gj} with a slightly more restrictive condition on $\om,$ Theorem \ref{thgj} holds as stated by
a result of \ci{bj3}. For the results obtained by methods stemming from the approach of \ci{bg} one has

\bt {\rm \ci{bj}} In Theorems \ref{thbg},\ref{strip},\ref{Thcdi},\ref{d=2} dynamical localization also holds. \et

This also applies to other results on nonperturbative localization, e.g. \ci{bgs2}.

\section {Quantum kicked rotor} \la{kicked}\setzero

\vskip-5mm \hspace{5mm}

The quantum kicked rotor was introduced in \ci{cis} as a model in quantum chaos. It is given by the time-dependent
Schrodinger equation on $L^2(\TT)$
$$
i\frac{\partial\phi}{\partial t} = a\frac{\partial^2\phi}{\partial^2\theta}+ib\frac{\partial\phi}{\partial \theta}+V(t,\theta)\phi
$$
where $V(t,\theta)=\kappa \cos 2\pi\theta \sum_{n\in \ZZ}\delta(t-n).$
It represents
quantization of the Chirikov standard map, and a conjecture (e.g.\ci{bel}) was that for a.e. $a,b$ the solution $\phi$ is
almost-periodic in time, thus demonstrating ``quantum suppression of chaos" (as some chaos is expected for the standard map). Such
almost periodicity follows from dynamical localization for the Floquet operator
$W:L^2(\TT)\to
L^2(\TT)$ defined by
$W\phi(t,\theta)=\phi(t+1,\theta).$ $W$ is a unitary operator that in Fourier representation can be written as a product $U\cdot S$
where $S(n,m)=S(n-m)$ is a T\"oeplitz operator with very fast decay of $S(n)$ and $D(n,m)=D(n)\delta(n,m)$ is a diagonal operator
with $D(n)=\exp(2\pi i(T^nx)_2),$ with $T$ being the skew shift of the torus $T(x_1,x_2)=(x_1+\om,x_2+x_1),$ and $x_1,x_2,\om$
determined by
$a,b.$ The nonperturbative methods (particularly, the method for skew-shift dynamics localization in \ci{bgs2} and a long-range
method \ci{bj2} ) were further developed for this model to obtain

\bt {\rm \ci{bkicked}} For any $\eps >0$ and any fixed $b$ there is $\kappa(\eps),$ and for $\kappa <
\kappa(\eps),$ $\Omega(\kappa) \subset \TT$ with $\mbox{mes} (\Omega) <\eps,$ s.t. for $a \notin \Omega (\kappa),$
operator $W$ has DL.
\et

Exploiting the multiplicative nature of $W$ one also obtains a nonperturbative counterpart:

\bt {\rm \ci{bj4}} There is $\kappa_0 >0$ such that for any $b,$ operator $W$ has dynamical localization for
$\kappa <\kappa_0$ and a.e. $a.$
\et

This confirms the ``quantum suppression of chaos" conjecture for small $\kappa.$

\label{lastpage}

\end{document}